\documentclass[aps,showpacs,floatfix,twocolumn]{revtex4}
\usepackage{graphicx}
\makeatletter
\newbox\slashbox \setbox\slashbox=\hbox{\large$/$}
\def\pslash#1{\setbox\@tempboxa=\hbox{$#1$}
\@tempdima=0.5\wd\slashbox \advance\@tempdima 0.5\wd\@tempboxa
\copy\slashbox \kern-\@tempdima \box\@tempboxa}

\makeatother
\newcommand{\mat}{\left ( \begin{array}{cc}}
\newcommand{\emat}{\end{array} \right )}
\newcommand{\be}{\begin{eqnarray}}
\newcommand{\ee}{\end{eqnarray}}
\newcommand{\ba}{\begin{array}}
\newcommand{\ea}{\end{array}}

\newcommand{\ben}{\begin{eqnarray*}}
\newcommand{\een}{\end{eqnarray*}}

\def\beq{\begin{equation}}
\def\eeq{\end{equation}}
\begin{document}
\title{Semi-Poisson statistics in quantum chaos}
\author{Antonio M. Garc\'{\i}a-Garc\'{\i}a}
\affiliation{Physics Department, Princeton University, Princeton, New Jersey 08544, USA}
\affiliation{The Abdus Salam International Centre for Theoretical
Physics, P.O.B. 586, 34100 Trieste, Italy}
\author{Jiao Wang}
\affiliation{Temasek Laboratories, National University of Singapore,119260 Singapore.}
\begin{abstract}
We investigate the quantum properties of a non-random Hamiltonian with 
a step-like singularity. It is shown that the eigenfunctions are multifractals and, 
in a certain range of parameters, the level statistics is described exactly by Semi-Poisson 
statistics (SP) typical of pseudo-integrable systems. It is also shown that our results
are universal, namely, they depend exclusively on the presence of the step-like 
singularity and are not modified by smooth perturbations of the potential or the
addition of a magnetic flux.
%A1
% which breaks the original time reversal invariance of 
%the system. 
Although the quantum properties of our system are similar
to those of a disordered conductor at the Anderson transition,
%A1 
%(scale invariant spectrum,
%level repulsion, sub-Poisson number variance and multifractal wavefunctions), 
we report important
quantitative differences in both the level statistics and the multifractal 
dimensions controlling the transition. Finally the study of quantum transport 
properties suggests that the classical singularity induces quantum anomalous 
diffusion. We discuss how these findings may be experimentally corroborated by 
using ultra cold atoms techniques.
\end{abstract}
\maketitle
%A1 New 1st par. in order not to overlap with my recent paper. Sorry!. 
It is by now well established that the
 analysis of the level statistics is one of the main tools
 in the study of quantum complex systems. Moreover
the spectrum, unlike the wavefunctions, is easily
accessible either numerically or experimentally. 
Part of this interest stems from the fact that, once the model-dependent
spectral density is extracted from the spectrum, 
the level correlations of apparently unrelated 
models shows striking universal features in a  variety of physical situations.
For instance, the celebrated 
Bohigas-Giannoni-Schmit conjecture (BGS) \cite{oriol} states the level statistics 
of a quantum system whose classical counterpart is deterministic but fully chaotic 
does not depend on the microscopic details of the Hamiltonian but only on the 
global symmetries of the system 
and coincides with those of a random matrix
with the same symmetry (usually referred to as Wigner-Dyson statistics (WD)\cite{mehta}).

Remarkably the same WD also describes \cite{efetov} the  spectral 
correlations of a disordered  system \cite{anderson} 
in the metallic limit. By contrast, for 
 disorder strong enough, eigenstates localization becomes important, 
the spectrum is not correlated and the 
level statistics is universally described by Poisson statistics. 
For deterministic systems
Poisson statistics appears provided that the classical dynamics is 
integrable \cite{tabor}.

%A remark is in order, unlike the spectrum, the eigenfunction properties 
%may depend on the representation utilized to express the Hamiltonian. 
%Thus localization studies in disordered systems usually express the 
%Hamiltonian  in a real space representation but for deterministic systems
%is typical a momentum representation.
Despite its robustness, the universality associated to 
 WD has also 
clear limits of applicability. 
% For instance the spectral correlations of quantum deterministic chaotic
% systems are supossed %to be described by WD statistics only up 
%to a energy scale associated to the shortest periodic orbit $E_per$ 
%(semiclassically %the spectral density can be expressed only in terms of 
%the periodic orbits of the classical counterpart)Spectral correlations 
%among eigenvalues separated by energies larger than  $E_per$ will 
%involve  the specific details of the Hamiltonian and consequently will
%be no universal.
For instance, the quantum properties of Hamiltonians whose classical phase space 
is a superposition of chaotic and integrable parts are supposed to depend 
dramatically on the details of the Hamiltonian and consequently their 
properties are non-universal. Similarly, for finite disordered systems in the 
metallic regime, the dimensionless conductance $g = E_c/\Delta$ ($E_c$, the
Thouless energy, is a scale of energy associated with the classical diffusion
time through sample and $\Delta$ is the mean level spacing) sets the number
of eigenvalues whose spectral correlations are universally described by
WD. 
%Beyond $g$ the spectral correlations are not universal, they depend itself on $g$ 
%and other features as boundary conditions or the shape of the cavity.

Universality in the spectral correlations has also a counterpart in the 
eigenfunctions properties. Thus Poisson statistics is associated with 
exponential localization of the eigenfunctions and WD is typical 
of systems in which the eigenstates are delocalized through the sample and 
can be effectively represented by a superposition of plane waves with 
random phases. 

%A1
%In principle it is not an easy task to find other windows of universality 
%besides the ones described by WD and Poisson statistics, and if they exist 
%so what is its extent and how they are defined in terms of eigenfunctions 
%and level statistics.
Recently it was reported \cite{sko,kravtsov97,chi} that the quantum 
properties of a disordered conductor at the metal-insulator 
 transition (usually referred to as the Anderson transition (AT)) are to certain 
 extent also 
 universal. 
%This transition 
%is observed in more than two 
%dimension around the center of the band for a critical amount of disorder.    
Features related to this new universality class include
 multifractal eigenstates \cite{aoki}
and level statistics given by critical statistics \cite{kravtsov97}(see below for a 
definition). 
Intuitively \cite{kra96} multifractality means that eigenstates have 
structures at all scales. Roughly speaking the amplitude of probability of 
a multifractal eigenstate has peaks (`probability splashes') at all scales 
decaying as a power-law from its maximum. 
Consider the volume of the subset
of a box for which the absolute value of the wave function $\Psi$ is larger
than a fixed number $M$. If this volume scales as $L^{d^*}$  (with $d^*< d$),
then $d^*$ is called the fractal dimension of $\Psi$. In case the 
fractal dimension depends on the value of $M$, the wave function is said to
be multifractal. On a more formal level multifractality is defined 
either through the box counting method (see \cite{cuevas} and the eigenvector analysis 
 section below) or 
the anomalous scaling of the eigenfunction moments 
${\cal P}_q=\int d^dr |\psi({\bf r})|^{2q}$ with respect to the sample size
$L$ as ${\cal P}_q\propto L^{-D_q(q-1)}$, where $D_q$ is a set of exponents
describing the AT \cite{aoki}. 
%The multifractal nature of the eigenfunctions is also the origin of the 
%anomalous quantum diffusion observed at the AT. 
`Critical statistics' \cite{sko,kravtsov97} (the level statistics at the AT)
is intermediate between Wigner-Dyson and Poisson statistics. Typical features
include: scale invariant spectrum \cite{sko}, level repulsion and asymptotically 
linear number variance \cite{chi}. 
%A1

Level and eigenfunction correlations at the AT are said to be universal in the sense that
typical features of critical statistics as level repulsion, scale invariance and
linear number variance
($\Sigma^{2}(L)=\langle L^2 \rangle - {\langle L \rangle}^2 = \chi L$ for $\chi < 1$ and $L \gg 1$)
do not depend on boundary
conditions, the shape of the system or the microscopic details of the disordered potential \cite{montam}.
However the slope of the number variance or the 
functional form of certain level correlators as the level spacing distribution $P(s)$ (the probability of
 having two eigenvalues at a distance $s$) may 
 depend on additional parameters as the dimensionality of the space.

Another argument reinforcing the universality of critical statistics is the fact
that, as for WD, which describes the level statistics of a Gaussian 
random matrix model, critical statistics has also been found in a variety of 
generalized random matrix models: based on soft confining potentials \cite{log}, 
effective eigenvalue distributions \cite{Moshe,ant4} related to the 
Calogero-Sutherland model at finite temperature and random banded matrices with 
power-law decay \cite{ever}. The latter is specially interesting since an AT has
been analytically established by mapping the problem onto a non linear $\sigma$ model.

Finally we recall that critical statistics is not related to any 
ergodic limit of the quantum motion as in the case of WD.
Consequently it is capable to describes non trivial dynamical 
features and its limit of validity is not restricted to a scale given
by the dimensionless conductance $g$ (as for WD). 
%which at the AT is about the unity.

%In a certain region of parameters all of these models share the same spectral kernel,
% \be
%\label{ow
% \bar K_T(x,0)=T\frac{\sin(\pi x)}{\sinh(\pi xT)}
%\ee
%$K(s)=T\frac{\sin(\pi s)}{\sinh(\pi sT)}$ ($T$ is a free parameter which enters 
%in the definition of the above models) which often is taken as a definition 
%of critical statistics.

A natural question to ask is whether 
critical statistical and multifractal wavefunctions are exclusive of disordered 
 systems or may
also appear in deterministic quantum systems. Indeed, in a recent letter \cite{ant9}
we have established a novel relation between the presence of anomalous diffusion in 
the classical dynamics, the singularities of a classically chaotic potential and the
power-law localization of the quantum eigenstates. Specifically, for a certain kind 
of singularity ($\log$ for 1+1D system) associated to classical $1/f$ noise, it is
found that the level statistics is described by critical statistics and the eigenstates
are multifractal with a $D_q$ quantitatively similar to the one at the AT. These results 
are universal in the sense that neither the classical nor the quantum properties 
depend on the details of the potential but only on the type of singularity.

Other non-random systems 
 whose level statistics has been reported to 
 be similar to critical statistics include: 
Coulomb billiards \cite{altshu}, Anisotropic Kepler problem 
\cite{wintgen}, generalized kicked rotors \cite{bao} and pseudointegrable billiards 
\cite{bogo3,bogo04}. For the latter the dynamics is intermediate
between chaotic and integrable
 and, in order to fit the spectrum, Bogomolny and coworkers
introduced \cite{bogo3} a purely phenomenological short range plasma model whose 
joint distribution of eigenvalues  {\cite{bogo3}} is given by the classical Dyson
gas with the logarithmic pairwise interaction restricted to a finite number $k$ 
of nearest neighbors. Explicit analytical solutions are available for general $k$.  
For instance for $k=2$ (usually referred to as Semi-Poisson statistics SP), 
 $R_2(s)=1-e^{-4s}$, $P(s)=4s e^{-2s}$ and 
$\Sigma^2(L) = L/2 +(1-e^{-4L})/8$ where $R_2(s)$ is the two level 
correlation function (TLCF). 
In passing we mention that SP can also be obtained by 
removing every $k$ eigenvalue out of a spectrum with Poisson statistics. 
It turns out that this short-range plasma model reproduces typical characteristics 
of critical statistics as level repulsion and linear number variance with a slope 
depending on $k$. 

However SP is quantitatively different from 
critical statistics. In critical statistics, as mentioned above, the  joint distribution of eigenvalues 
can be considered as an ensemble of free particles at finite temperature with 
a nontrivial statistical interaction. The statistical interaction resembles the
Vandermonde determinant and the effect of a finite temperature is to suppress the
correlations of distant eigenvalues. In the case of SP this suppression is
 abrupt \cite{bogo3} since only nearest neighbor levels can interact.
Thus critical statistics and SP share similar 
generic features but are in principle quantitatively different.

%A1
 The aim of the letter is to
establish under what generic circumstances one may expect SP in 
the context of non-random Hamiltonians.

For the sake of clearness we enunciate our main conclusions:
We have found that the appearance of SP can indeed be traced back to the 
presence of a certain kind of singularity (different from the one associated 
to critical statistics) in the classical potential. It is shown that SP 
is indeed robust under arbitrary smooth perturbations of the classical potential or the 
 insertion of a magnetic flux provided 
that the classical singularity is preserved. The eigenfunctions associated to 
SP are found to be multifractal but quantitatively different from those of
 a disordered conductor at the AT. Finally we argue that quantum anomalous diffusion induced by the 
 classical singularity may be verified experimentally by using ultra-cold atoms techniques.

 The organization of the paper is as follows, in the
 next section we introduce the model: a generalized kicked rotor in a potential 
with a step-like singularity. In section two we discuss analytical results available
for our model, then we investigate the level statistics and perform a multifractal analysis 
of the eigenstates. Finally, in section three 
we examine the quantum diffusion of our model and discuss 
possible ways of experimental verification.

\section{The model} 
We investigate a generalized kicked rotor in $1+1$D with a step-like singularity,
\be
\label{ourmodel}
{\cal H}= \frac{p^2}2 +V(q)\sum_n\delta(t -nT)
\ee
with $q \in [-\pi,\pi)$, $V(q)$ is an arbitrary nonanalytical function with a
step-like singularity. The simplest case corresponds to, 
\be
\label{potential} 
V(q) = \cases{v_0,&if $q\in [-a,a)$\cr 0,&otherwise.\cr}
\ee 
where $a$ sets the size of the step and $v_0$ the height. We shall see in a broad 
range of parameters our results do not depend on the specific form of $V(q)$ but
only on the presence of the singularity.

The quantum dynamics is governed by the quantum evolution operator $\cal U$ over
a period $T$. Thus, after a period $T$, an initial state $\psi_0$ evolves to 
$\psi(T) = {\cal U}\psi_0 = e^{\frac{-i {\hat p}^2T}{4{\bar h}}}
e^{-\frac{iV(\hat q)}{\bar h}}e^{\frac{-i {\hat p}^2 T}{4{\bar h}}}\psi_0$ 
where $\hat p$ and $\hat q$ stands for the usual momentum and position operator. 
Our aim is to solve the eigenvalue problem 
${\cal U}\Psi_{n}=e^{-i\kappa_n/ \hbar}\Psi_{n}$ where $\Psi_{n}$ is an 
eigenstate of $\cal U$ with quasi-eigenenergy $\kappa_n$. In order to proceed we 
first express the evolution operator $\langle m| {\cal U}
| n \rangle = U_{nm}$ in a basis of momentum eigenstates $\{| n \rangle = 
\frac {e^{in \theta}}{\sqrt{2\pi}}\}$ with $n = 0, \ldots N \rightarrow \infty$, 
\be
\label{uni0}
U_{mn} = \frac{e^{-i \frac{T \hbar}{4} (m^2+n^2)}}{2\pi} 
\int_{-\pi}^{\pi}dq e^{iq(m-n)-i V(q)/\hbar}
\ee
We remark that in this representation, referred to as `cylinder representation',
the resulting matrix  $U_{nm}$ is unitary exclusively in the $N \rightarrow \infty$
limit. For practical calculations this is certainly a disadvantage since
 besides typical
finite size one has also to face truncation effects, namely, 
the integral of the density of
 probability is not exactly the unity and eigenvalues are not 
pure phases ($e^{-i\theta_n}$) as expected in a Unitary matrix.

An alternative procedure, referred to as `torus representation', specially suited 
for numerical calculations, is to make the Hilbert space finite 
(i.e.$~m,n=1,\ldots N$) but still keep the matrix Unitary. This can be achieved 
by imposing periodicity (with period $P \sim N$) in momentum space. 
With these choices $\hbar$ is set to the unity since the period 
in momentum $P \sim N$ and the period in position $Q \sim 2\pi$ are related by $PQ= 2\pi N\hbar$.
Thus the period on momentum is large ($P=N$) 
and the phase space effectively resembles a cylinder (for more 
details see \cite{reviz}). 
 
In order to keep the kinetic term of the evolution 
matrix also periodic, we take $T=2\pi M/N$ with $M$ an integer (not a divisor 
of $N$). For technical reasons (see below) $M$ is chosen to make $T$ roughly
constant ($\approx0.1$ in this paper) such that $M/N$ be a good approximation 
to a irrational for every $N$ used. 
The resulting evolution matrix (for $N$ odd) then reads
\be
\label{uni}
\langle m| {\cal U}| n \rangle = \frac{1}{N}e^{-i \frac{\pi M}{2N}{(m^2+n^2)}}
\sum_{l}e^{i\phi(l,m,n)}
\ee
where $\phi(l,m,n)= 2\pi (l+\theta_0)(m-n)/N-V(2\pi (l+\theta_0)/N)$,
%$m,n = 1,\ldots N$,
$l = -(N-1)/2,\ldots (N-1)/2$ and $0 \le \theta_0 \le 1$; $\theta_0 $ 
is a parameter depending on the boundary conditions ($\theta_0=0$ for 
periodic boundary conditions). 
%The only disadvantege of this representation 
%is that one is adding a additional periodicity in the system which is not 
% present in the original Hamimltonian. However by setting the period
%proportional of the size system the difference  

In this paper we shall use both representations depending on the 
issue to be discussed. Thus for the analytical analysis
the cylinder representation is more appropriate: the limit $N \rightarrow \infty$
can be effectively taken and consequently truncation effects are absent. For 
the numerical calculations we use the torus representation due to the difficulty
to deal with truncation effects. An exception is the case of quantum diffusion 
where these effects can be accurately detected and subtracted from the calculation.
As a final remark we mention that the numerical evaluation of the eigenvalues and 
eigenvectors of $\cal U$ (either for the torus or the cylinder) is carried out
by using standard diagonalization techniques for volumes ranging from $N=500$ 
to $N=8000$. For $\theta_0=0$, parity is a good quantum number and consequently 
states with different parity must be treated separately.

\section{Results}

For the sake of clearness we first summarize our main results:

1. The level statistics associated to the 
  evolution matrix of the Hamiltonian Eq.(\ref{ourmodel}) is scale invariant and given 
  by SP in the region $S = \tan (v_0/2) \gg 1$ where
  $v_0$ is the height of the step-like potential. These results are 
 universal: they neither depend on the specific form of $V(q)$ (provided that the 
 singularity is preserved) nor on any source (as a magnetic flux) of 
time reversal symmetry.

2. In the limit $S = \tan (v_0/2) \ll 1$ the spectrum 
  is scale invariant and well described by Poisson statistics. 
  For intermediate values of $S$ we observe a transition from SP 
 to Poisson as $S$ is decreased. 

3. In the SP region the eigenstates are multifractal 
 but with a multifractal spectrum $D_q \sim {A}/{q}+D_{\infty}$ essentially different 
  from the one observed at the AT. 

4. The classical singularity induces quantum anomalous diffusion.

\subsection{Analytical results}
In this section we investigate what kind of analytical information can be obtained
from the Hamiltonian Eq.(\ref{ourmodel}). 
%First we will restrict
%ourselves to the simplest potential Eq. (\ref{potential}) though later on 
%we will show our findings do not depend on the specific form of the potential provided that
%the singularity is kept. 
Our aim is to show that in a certain range of parameter 
 the level statistics associated to Eq. (\ref{ourmodel}) is exactly given 
 by SP. In a first step we map, following the 
 standard prescription of Ref. \cite{fishman}, the evolution matrix Eq.(\ref{uni}) onto a 1D Anderson model.
For smooth potentials such mapping results in a standard 
1D Anderson model with short range hopping. As known, for this kind of models eigenstates 
are exponentially localized for any amount of disorder and consequently 
dynamical localization has been reported \cite{fishman} in the associated kicked-rotor system.
 
However in our case the situation is different. The classical 
step-like singularity induces long-range correlations in the associated 1D Anderson 
model, 
\be
\label{ourmodel1}
{\cal H} \psi_n= \epsilon_n \psi_n + \sum_m F(m-n)\psi_m
\ee
where $\epsilon_n \sim \tan(T n^2)$, 
\ben
F(m-n) &=&\int_{-\pi}^{-\pi}d\theta \tan(V(\theta)/2)e^{-i\theta(m-n)}\\
&=&A\frac{\sin \gamma (m-n)}{m-n}
\een
$A \sim \tan (v_0/2)$ and $\gamma = a $. 
For $\epsilon_n$ a random number from a box distribution
 $[-W/2,W2]$, it has been shown recently \cite{antonio10} (see discussion below)
without invoking any ensemble average that the level 
statistics is exactly given by SP in the limit $A \gg W$.
However as $A$ gets comparable to $W$ a shift to Poisson is observed.  Thus for 
our Hamiltonian Eq.(\ref{ourmodel}) the region in which SP holds corresponds with
$v_0 \sim \pi$ ($A = \tan (v_0/2) \gg 1$). Moreover $T$ must be such that 
$\epsilon_n \sim \tan(T n^2)$ is approximately random, this always occurs for $T \sim 1$ 
irrational. In the torus representation of the evolution matrix $T$
must be a rational so in order 
 to overcome this problem we set $T$ to be a good approximation to an irrational number.\\
For the sake of completeness the rest of this 
 section is devoted, following Ref.\cite{antonio10}, to a technical account
 of the reasons for the appearance of SP in the Hamiltonian Eq. (\ref{ourmodel1}). 
We recall that throughout the demonstration the diagonal 
 energy $\epsilon_n$ is assumed to be strictly random.
First of all we express the  Hamiltonian (\ref{ourmodel1}) in Fourier space as,
\ben
{\cal H} =  E_k\psi_k + \sum_{k \neq k'}{\hat A}(k,k')\psi_k'
\een
where 
\ben
E_k = \sum_r \frac{\sin \gamma r}{r} e^{i k r} 
\een
and 
\ben
{\hat A}(k,k') =\frac{1}{N}
\sum_n \epsilon_n e^{-in(k-k')}.
\een
We fix $\gamma = \pi/2$ (our findings do not depend on 
 $\gamma$). After a simple calculation we found that $E_k$ is not a smooth function (this step-like singularity
 is indeed the seed for the appearance of SP), 
$E_k = A\pi/2$ for $k < \pi$ and $E_k = -A\pi/2$ for $ k > \pi$. 
 There are thus only two possible values of the  energy separated by a gap $\delta = A\pi$. 
  Upon adding a weak ($A \gg W$) disordered potential this degeneracy is lifted 
 and the spectrum is composed  of two separate bands
 of size $\sim W$ around each of the bare points $-A\pi/2,A\pi/2$.  Since the Hamiltonian 
 is invariant under the transformation $A \rightarrow -A$, the spectrum 
must also posses that symmetry. 
 That means that, to leading order in $A$ (neglecting $1/A$ corrections), 
the number of independent eigenvalues of Eq. \ref{ourmodel1} is $n/2$
 instead of $n$.

We now show how this degeneracy affects the roots (eigenvalues) of the characteristic polynomial $P(t)= \det (H-tI)$. 
Let 
\ben
P_{dis}(t)=a_0+a_1t+\ldots a_n t^n 
\een
 be the characteristic polynomial associated 
with the disordered part of the Hamiltonian. We remark that despite of its complicated form, its roots, by definition, are random numbers with a box distribution $[-W,W]$. On the other hand, in the clean case,
 \ben
P_{clean}(t)=(t-A)^{n/2}(t+A)^{n/2}
\een 
($\pi$ factors are not considered). 
Due to the $A \rightarrow -A$ symmetry, the 
full (Eq. \ref{ourmodel1}) case $P_{full}$ corresponds with $P_{dis}$ but replacing $t^k$ factors by a 
combination  $(t-A)^{k_1}(t+A)^{k_2}$ with $k_1 + k_2 = k$. The roots of $P_{full}$ will be 
in general complicated functions of $A$. However in the limit  of interest, $A \gg W \rightarrow \infty$, an analytical evaluation is possible. 
By setting $t = t_1 -A$ we look for roots $t_1$ of order the unity in the $A$ band.
We next perform an expansion of the characteristics polynomial $P_{full}$ to leading order in $A$.
 Thus we keep terms $A^{n/2}$ and neglect lower powers in $A$. The resulting $P_{full}$ is given by,
\ben
 P_{full}  = t_{1}^{n/2} +\frac{a_{n-2}{t_1}^{n/2-2}}{3}  
+\frac{a_{n-3}{t_1}^{n/2-3}}{4} & \\
+ \ldots 
2 \frac{a_{n/2+1}{t_1}}{n}+ 2\frac{a_{n/2}}{n+2}
\een
 where the coefficients $a_n$  are the {\it same} than those of $P_{dia}$ above but only $n/2$ of them appear in the full case. 
The eigenvalues $\epsilon'_{i}$ of Eq.({\ref{ourmodel1}}) around the $A$ band are $\epsilon'_i = A + \beta_i$ with $\beta_i$ a root of $P_{full}$. The effect of the long range interaction is just to 
 remove all the terms with coefficients $a_0$ to $a_{n/2}$ from the characteristic polynomial of the diagonal disordered case. The spectrum is thus that of a pure diagonal disorder 
where half of the eigenvalues have been removed. The remaining eigenvalues are 
 still symmetrically distributed (the 
  ones with largest modulus are well approximated by $t_{max}= \pm \sqrt{\frac{a_{n-2}}{3}}$) around $A$. 
That means, by symmetry considerations, that the removed ones must be either the odd or the even ones. 
This is precisely the definition of semi-Poisson statistics.
In conclusion, our model reproduces exactly the mechanism which is utilized in the very definition
 of SP. We finally mention that the only effect of the coefficients $3,4 \ldots, n+2/2$ 
is to renormalize the effective size of the spectrum: $\sim 2W$ for diagonal disorder and $\sim 2W/\sqrt{3}$
 for the Eq. (\ref{ourmodel1}). 

\subsection{Level statistics}

The above analytical arguments have been fully corroborated by numerical calculations. 
As mentioned previously in order to avoid truncation effects we have chosen the torus 
representation of the evolution matrix.
%%%%%%%%%%%%Fig 1%%%%%%%%%%%%%%%%%%%%%%%%%%%%%%%%%%%%%%%%%%%%%%%%%%%%%%%%%%%%%%%%%%%%
\begin{figure}
\includegraphics[width=.95\columnwidth,clip]{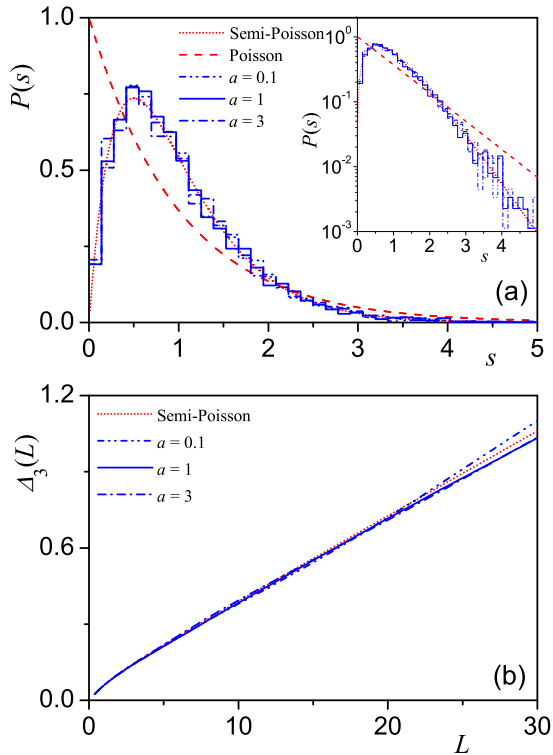}
\caption{(Color online) Level spacing distribution $P(s)$ (a) and spectral rigidity $\Delta_3(L)$ 
(b) for the spectrum of the evolution matrix Eq.(\ref{uni}) with potential Eq.(\ref{potential}) 
for different widths $a$ and $v_0 = \pi$. 
In all cases the size of the evolution matrix is $N=6397$. The agreement with SP is excellent.}  
\label{fig1}
\end{figure}
%%%%%%%%%%%%%Fig 1%%%%%%%%%%%%%%%%%%%%%%%%%%%%%%%%%%%%%%%%%%%%%%%%%%%%%%%%%%%%%%%%%%%%
Our first goal is to show that the level statistics of our model with the potential
of Eq. (\ref{potential}) is given by SP in the region  $S = \tan (v_0/2) \gg 1$ for any
$a$. In Fig. 1 we plot the level spacing distribution $P(s)$
 for different values of $a$ and $S\gg1$. As observed,
the agreement with SP is excellent in all cases even in the tail of the distribution
and it is not restricted to short range correlators.
As shown in Fig. 1b, the study of long range correlators as the spectral rigidity 
$\Delta_3(L)=\frac{2}{L^4} \int_0^L (L^3-2L^2r+r^3)\Sigma^2 (r) dr$
further confirms this point. Up to scales of $30-40$ eigenvalues 
deviations from SP are almost indistinguishable
 for different values of the parameters. Deviations for larger scales
are due to well known finite size effects. 
Although not shown, we have also checked that in the range of volumes accessible to 
numerical techniques, the level statistics was to great extent scale invariant, namely, 
it was independent from the system size.
%%%%%%%%%%%%Fig 2%%%%%%%%%%%%%%%%%%%%%%%%%%%%%%%%%%%%%%%%%%%%%%%%%%%%%%%%%%%%%%%%%%%%
\begin{figure}
\includegraphics[width=.95\columnwidth,clip]{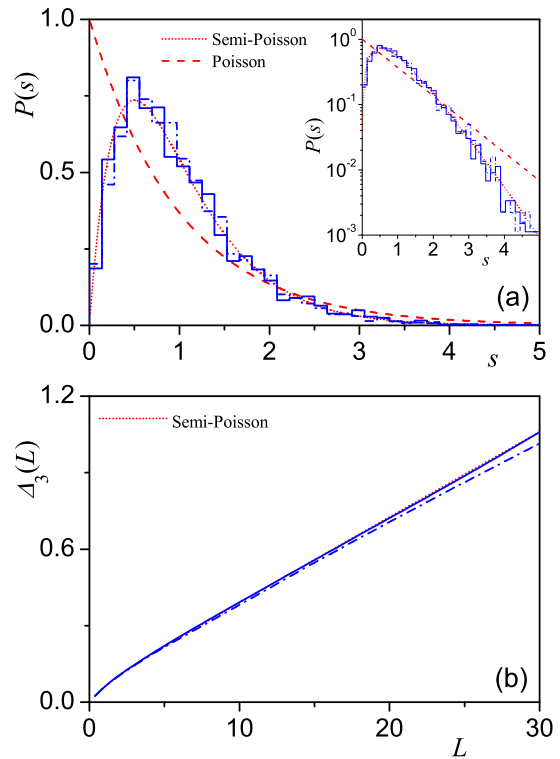}
\caption{(Color online) Level spacing distribution $P(s)$ (a) and spectral rigidity $\Delta_3(L)$
(b) of the spectrum of the evolution matrix ($N=6397$) Eq.(\ref{uni})
 with potential Eq.(\ref{potential}) and $v_0=\pi$ and $a=\pi/2$.
As observed the level statistics is not affected
 neither by the breaking of the time reversal invariance
(dashed-dotted) line nor by the addition of a smooth perturbation (solid line)
$V_{per}=c_1 \cos(q)+c_2 \sin(q)+c_3 \cos(2q)+c_4 \sin(2q)$ with random $c_i$ (see text).}
\label{fig2}
\end{figure}
%%%%%%%%%%%%%Fig 2%%%%%%%%%%%%%%%%%%%%%%%%%%%%%%%%%%%%%%%%%%%%%%%%%%%%%%%%%%%%%%%%%%%%
Once the region in which SP holds has been established, we
 investigate the robustness/universality of these results. In order to proceed we 
have repeated the level statistics analysis for a potential with a non-analytical 
step-like form but perturbed by a smooth chaotic potential. As shown in Fig.2, neither
short nor long range spectral correlators are affected by the chaotic perturbation 
provided that the step-like non-analyticity is preserved. The perturbation potential
$V_{per}$ used in Fig. 2 is defined by $V_{per}=c_1 \cos(q)+
c_2 \sin(q)+c_3 \cos(2q)+c_4 \sin(2q)$ with $c_i$ randomly chosen from a uniform 
distribution $(0,1)$. We have also checked that higher frequency components $\sim \cos(3q)$
 do not change the results provided that $|V_{pert}| \ll v_0$.

In addition, we have also found that the level statistics is not modified if time 
reversal invariance is broken by adding a magnetic flux to Eq. (\ref{ourmodel})
(which is equivalent to setting $T$ an irrational multiple of $2\pi$; see \cite{ant9}). 
These results indicate that SP is universal; namely, it does not depend on the details 
of the potential provided that the step-like singularity is still present.

Finally we study the transition to Poisson statistics as the parameter $S = \tan (v_0/2)$ 
goes from $S \gg 1$ (SP region) to $S \ll 1$. As shown in Fig.3 the level statistics 
(both $P(s)$ and $\Delta_3(L)$) seems to move smoothly from SP to Poisson. All of the 
typical features of criticality above mentioned are maintained through the transition.
Of course parameters as the slope of the number variance $\chi$ runs from $\chi = 1/2$ 
(SP) to $\chi = 1$ (Poisson) as $S$ is decreased. However, we stress that, 
unlike the SP region, it is hard to unambiguously define a universality class in the
transition region. The point is that, in essence, the appearance of SP can be traced
back (see analytical analysis) to the gap in the spectrum of the Hamiltonian.
The step-like singularity thus separates the spectrum in two different 
bands around the only two eigenvalues in the clean case. As 
the disorder strength becomes comparable with the numerical value of the 
 bare eigenvalues both bands mix up and the spectrum deviates from
SP. We could not find any sign of universality  
in the way in which this mixing occurs since it may depend on the microscopic details of the
potential. 
%A1
%By putting together the two bands the sharp nearest neighbor interaction 
%induced by the singularity is gradually washed up by the effect of increasing disorder. 
Due to this lack of universality in the rest of the paper we will focus 
our investigation mainly on the region of parameters associated
to SP. 
%%%%%%%%%%%%Fig 3%%%%%%%%%%%%%%%%%%%%%%%%%%%%%%%%%%%%%%%%%%%%%%%%%%%%%%%%%%%%%%%%%%%%
\begin{figure}
\includegraphics[width=.95\columnwidth,clip]{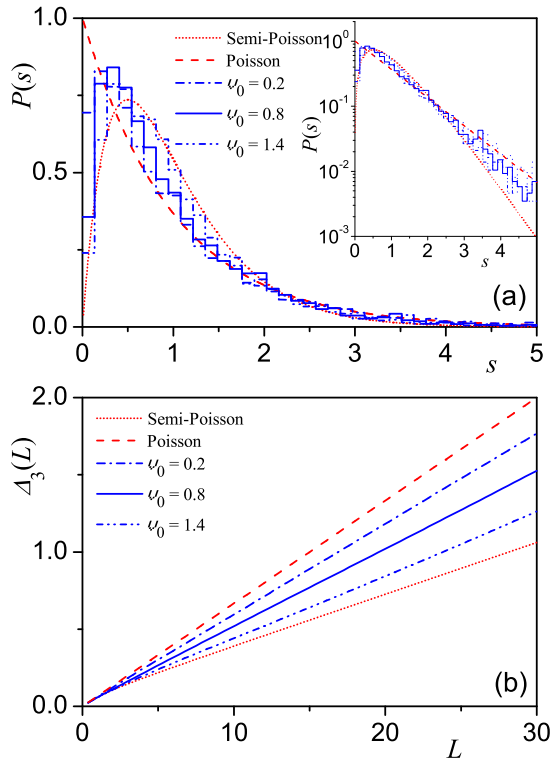}
\vspace{-.7cm}
\caption{(Color online) Level spacing distribution (a) and spectral rigidity (b) of the evolution matrix ($N=6397$) Eq.(\ref{uni})
for the potential Eq.\ref{potential} for different $v_0$ and $a = 1$. A transition from 
SP to Poisson is found as $v_0$ goes from $\pi$ to $0$.}
\label{fig3}
\end{figure}
%%%%%%%%%%%%%Fig3 %%%%%%%%%%%%%%%%%%%%%%%%%%%%%%%%%%%%%%%%%%%%%%%%%%%%%%%%%%%%%%%%%%%%
\subsection{Eigenvector analysis}   
We now investigate the eigenvector properties of the Hamiltonian Eq. (\ref{ourmodel})
with potential Eq. (\ref{potential}) in the region $S \gg 1$ corresponding
to SP.

This choice stems from the fact that, as shown above, 
the quantum properties do not depend on the details $V(q)$ but only on the presence of 
the step-like singularity.  For the numerical calculation we have again utilized the torus
representation of the evolution matrix in order to avoid leaking of probability due 
to truncation effects. 

%%%%%%%%%%%%Fig 4%%%%%%%%%%%%%%%%%%%%%%%%%%%%%%%%%%%%%%%%%%%%%%%%%%%%%%%%%%%%%%%%%%%%
\begin{figure}
\includegraphics[width=.95\columnwidth,clip]{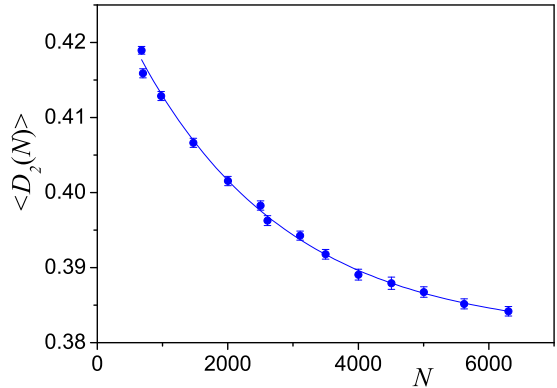}
\vspace{-.7cm}
\caption{(Color online) System size dependence of the multifractal dimension $\langle D_2(N) \rangle$.
 For each $N$ the average is taken over around 
$36000$ eigenstates of the evolution operator Eq.(\ref{uni}) 
 with the potential given by Eq. (\ref{potential}) and $a=1$, $v_0=\pi$. 
The best fitting function is a 
 exponential $(0.0493\pm 0.0007)e^{-N/(2290\pm 122)}+D_2$ with 
$D_2=0.381\pm 0.001$ (solid curve).}
\label{figure4}
\end{figure}
%%%%%%%%%%%%%Fig 4%%%%%%%%%%%%%%%%%%%%%%%%%%%%%%%%%%%%%%%%%%%%%%%%%%%%%%%%%%%%%%%%%%%%

We have two clear objectives in this section. On the one hand we would like to investigate 
whether the eigenvectors associated to SP are multifractal as at the AT. Once this question
is answered affirmatively our intention is to provide a careful and detailed analysis of 
the anomalous dimension $D_q$
%( ${\cal P}_q=\int d^dr |\psi({\bf r})|^{2q}$ with respect to the sample size $L$ as 
%${\cal P}_q\propto L^{-D_q(q-1)}$)
controlling the eigenstates multifractality. Based on the numerical findings we conjecture
the relation  $D_q = A/q+D_{\infty}$ which we claim
to be valid for all systems with SP.  \\
To start with we give a detailed account on how $D_q$ was computed. We use the standard
box-counting procedure \cite{cuevas}. In doing so, for a given eigenvector 
$\psi=\sum_{k=1}^{N}\psi_k|k \rangle$ of the system we first distribute all the components
into $N_l=N/l$ boxes of the same size $l$, then associate a probability  $p_i=\sum_{k}|\psi_k|^2$
 to each box  with $k\in [il,(i+1)l)$; $i=1,...,N_l$. The normalized 
$q$th moments of $p_i$ define the generalized fractal dimensions \cite{hent}:
\be
\label{Dq}
D_q(N)=\frac{1}{q-1}{\lim_{\delta \to 0}}\frac{\log \sum_{i=1}^{N_l}p_{i}^{q}}{\log\delta},
\ee
where $\delta=l/N$ is the size of the box normalized by the system size $N$. In practical
calculations $D_q$ is evaluated over an appropriate range of $\delta$ by performing
a linear regression of ${\log \sum_{i=1}^{N_l}p_{i}^{q}}$ with ${\log\delta}$ (usually 
$\alpha/N\le\delta\le 1/2$ where $\alpha$ is a characteristic microscopic length scale 
of the system \cite{cuevas}). In our case a good linear dependence of 
${\log \sum_{i=1}^{N_l}p_{i}^{q}}$ on ${\log\delta}$ is observed in the region 
$\delta\in (0.022,0.45)$ for a broad range of the potential parameters $v_0,a$ 
of Eq. (\ref{potential}). $D_q(N)$ is thus computed by linear regression over this 
$\delta$-window for each eigenvector.  
%A1
Since $D_q(N)$ is a self averaging \cite{ever}
 quantity the mean value $\langle D_q(N)\rangle$ over all eigenstates provides with 
 an meaningful description of the multifractals properties of the system. 
%%%%%%%%%%%%%Fig 5%%%%%%%%%%%%%%%%%%%%%%%%%%%%%%%%%%%%%%%%%%%%%%%%%%%%%%%%%%%%%%%%%%%%%
\begin{figure}
\includegraphics[width=.95\columnwidth,clip]{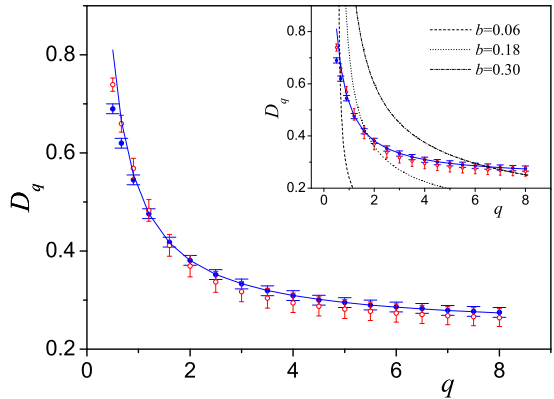}
%\vspace{-.7cm}
\caption{(Color online) Generalized fractal dimensions $D_q$ for the eigenvectors of the evolution  
 matrix Eq. (\ref{uni}) with potential Eq. (\ref{potential}) and 
$a=1$, $v_0=\pi$. Solid dots are the results from  Eq. (\ref{Dq})
and the open dots are those obtained from the scaling of the averaged 
$\log{\cal P}_q$ with $\log N$ (see 
 text). The solid curve is the conjecture $A/q+D_{\infty}$ with $A=0.286\pm0.001$ and 
$D_{\infty}=0.238\pm 0.001$. Inset shows the same results together with 
the prediction of critical statistics Eq. (\ref{DqAT}) for $b=0.06$, $ 0.18$ and $0.30$ respectively.}
\label{figure5}
\end{figure}
%%%%%%%%%%%%%Fig 5%%%%%%%%%%%%%%%%%%%%%%%%%%%%%%%%%%%%%%%%%%%%%%%%%%%%%%%%%%%%%%%%%%%%%
%A1 I rewrote this par.
We observe that, even after this averaging, the statistical fluctuations
of $\langle D_q(N)\rangle$ are still quite strong. 
In order to minimize these fluctuations we perform an additional average over 
realizations of the evolution operator Eq. ({\ref{uni}}) until we 
obtain around of $36000$ eigenvalues for each value of $N$ (see Fig. 4).
We generate different realization of the evolution matrix by
varying of the period $T$.
 In the time reversal invariant case ($T=2\pi M/N$)
this is done by picking up different values of $M$ and in the broken time reversal case
($T=2\pi \beta$ with $\beta$ irrational) by choosing different values of $\beta$.

As observed in Fig. 4,  
the $\langle D_q(N)\rangle$ thus obtained has a significant dependence on the system 
size which gets smaller as we approach the thermodynamic limit. In order to remove this
 finite size effect from $\langle D_q(N)\rangle$ we have tried different fittings,
$\langle D_q(N)\rangle = D_q +f(N,a,b)$ with $f(N,a,b) \rightarrow 0$ as $N \rightarrow \infty$
and $a,b,D_q$ fitting parameters. Thus the parameter $D_q$ 
 corresponds with the $N \rightarrow \infty$ limit of $\langle D_q(N)\rangle$. 
 We found that for various values of $q$
 the choice of $f$ that best describes the finite size corrections is $f(N,a,b) = ae^{-N/b}$
(see Fig. 4 for the case $q = 2$).

After the technical analysis we are ready to present our results. First of all we have
found that the eigenstates are indeed multifractal. As shown in Fig. 5, $D_q$ (the
$N \rightarrow \infty$ of $\langle D_q(N)\rangle $) depends clearly on $q$. 
We have also found that $D_q$ is very robust. Thus it is not affected by 
different choices of $a$, 
$v_0 \sim \pi$, smooth perturbations to the potential Eq. (\ref{potential}),
or by the breaking of the original time reversal symmetry.

The next task we face is try to find an explicit expression for $D_q$. Unfortunately 
we are not aware of any analytical technique capable of obtaining $D_q$ exactly. The 
comparatively short range of $q$ values for which numerical calculations are available
together with the statistical fluctuations make a numerical fitting ambiguous in the sense that 
many different fitting may yield quite satisfactory results. On the other hand we 
think it is extremely important to have a proper characterization of $D_q$ in order
to completely
define the universality class associated to SP. 
We found that the simplest expression for $D_q$ still compatible with the numerical
analysis is $D_q = A/q+D_{\infty}$ (with $A \sim 0.28$, and $D_{\infty}\sim 0.25$). 
As shown in Fig. 5 the agreement with the numerical 
results is very good. 
Remarkably, we have obtained a similar $D_q$ in the 
 study of the quantum evolution matrix associated to certain classical interval-exchange maps 
whose level statistics is exactly given by SP \cite{bogo04}. Based on this finding we 
suggest that the $D_q$ above describes the multifractal 
properties of all systems with SP.

As a double check we have also evaluated $D_q$ via the scaling of 
 eigenfunction moments ${\cal P}_q=\int d^dr |\psi({\bf r})|^{2q}\propto L^{-D_q(q-1)}$ 
with respect to the sample size $L=N$. 
In particular $D_q$ was obtained by linear regression
of $\langle \log{\cal P}_q \rangle$ with respect to $\log N$. As in the previous case, the
 average is taken over all eigenvectors and different realizations of the evolution and 
 finite size effects are removed by an appropriate fitting.   
 The $D_q$ obtained in this way is in complete agreement
  with the previous one within the numerical errors.

Finally we compare the above $D_q$ with the predictions of critical statistics. First
of all we remark that a comparison with a 3D (or 4D) disordered conductor at the AT 
is not entirely satisfactory since parameters as the slope of the number variance 
($\sim 0.27$ in the 3D case) defining the AT are different from the SP prediction ($1/2$).
We thus find more appropriate to compare the SP results with those of a random 
banded matrix with a power-law ($\sim 1/r$) decay and bandwidth $b$ \cite{ever} where           
the spectral statistics is given by critical statistics and the multifractal character
of the eigenstates has been established analytically. In the limit $b \ll 1$,
closer to SP, it is found that for $q > 1/2$
\be
\label{DqAT}
D_q = \frac{4b}{\sqrt{\pi}}\frac{\Gamma(q-1/2)}{\Gamma(q)}.
\ee  
As shown in Fig. 5 (inset), this function does not describe the multifractal 
dimensions associated to SP for any $b$. 
Thus we can conclude that despite their similarities 
critical statistics and SP belong to different universality class.

\section{Quantum anomalous diffusion and experimental verification}

%%%%%%%%%%%%%Fig 6%%%%%%%%%%%%%%%%%%%%%%%%%%%%%%%%%%%%%%%%%%%%%%%%%%%%%%%%%%%%%%%%%%%%%
\begin{figure}
\includegraphics[width=.95\columnwidth,clip]{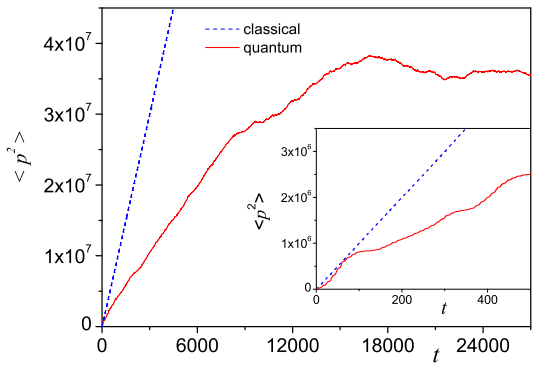}
\vspace{-.7cm}
\caption{(Color online) Classical and quantum variance $\langle p^2(t) \rangle$
of the density of probability $P(p,t)$ (see text)  
for the Hamiltonian Eq.(\ref{ourmodel}) ($T=1$,$\hbar=1$) 
and potential given by  Eq.(\ref{newmodel}) with $\sigma=10^{-4}$. The classical 
 counterpart was obtained from $10^6$ random initial conditions with zero
initial momentum and uniformly distributed $q$ in $(-\pi,\pi)$.}
\label{figure6}
\end{figure}
%%%%%%%%%%%%%Fig 6%%%%%%%%%%%%%%%%%%%%%%%%%%%%%%%%%%%%%%%%%%%%%%%%%%%%%%%%%%%%%%%%%%%%%

%%%%%%%%%%%%%Fig 7%%%%%%%%%%%%%%%%%%%%%%%%%%%%%%%%%%%%%%%%%%%%%%%%%%%%%%%%%%%%%%%%%%%%%
\begin{figure}[ht]
\includegraphics[width=.95\columnwidth]{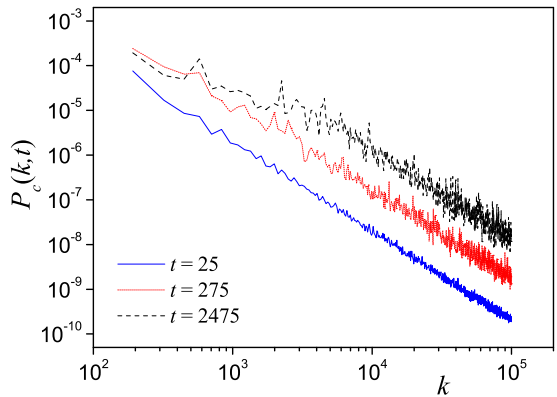}
\vspace{-.7cm}
\caption{(Color online) Quantum coarse-grained $P(k,t)$ versus $k$ at various times $t$ associated 
with the potential Eq. (\ref{potential}) ($v_0=\pi$ and $a=\pi/2$) plus a perturbation
$V_{per}(q)=8\cos(q)$. For a given pair of $k$ and $t$, $P_c(k,t)$ is the average of 
$P(k',t')$ over $k'=k-63,...,k+64$ and $t'=t-24,...,t+25$. $P_c(k,t)$ 
has power law tails $\sim t^{0.84 \pm 0.01} |k|^{-2}$ for $|k|\gg t/\hbar\rho$ where $\rho$ is the 
 spectral density . Similar 
results are observed if only the potential Eq. (\ref{potential}) is considered.}
\label{figure7}
\end{figure}
%%%%%%%%%%%%%Fig 7%%%%%%%%%%%%%%%%%%%%%%%%%%%%%%%%%%%%%%%%%%%%%%%%%%%%%%%%%%%%%%%%%%%%%
  
In this section we investigate how quantum transport properties are affected by the 
step-like non-analytical potential. Our motivation is twofold: On the one hand we 
wish to examine possible ways to test our results experimentally. On the other hand 
we are interested in finding out to what degree the phenomenon of dynamical localization typical of 
kicked rotors with smooth potentials $\sim \cos (q)$ is affected by the classical
singularity. We recall that dynamical localization for smooth chaotic potentials
manifests itself in the quantum suppression of classical diffusion in momentum space 
due to interference effects. Thus, contrary to the BGS conjecture \cite{oriol}, though the classical 
dynamics is chaotic eigenstates are exponentially localized in momentum space and 
the level statistics is Poisson.

We start by computing both the quantum and classical density of probability 
$P(k,t)=|\langle k|\phi(t)\rangle |^2$ of finding a particle with momentum 
$p = k\hbar$ at time $t$ for a given initial state $|\phi(0)\rangle=|0\rangle$ (initial condition
 in the classical case).

The first problem we face is that the potential Eq.(\ref{potential})
has a quite peculiar classical limit. The force exerted on the 
particle is a sum of two  Dirac delta functions $\delta(q\pm a)$, namely, it vanishes
elsewhere except at $q = \pm a$ where formally it diverges. 
Consequently the classical motion is that of a free
particle in a ring except for $q = \pm a$ where it gets an infinite force. 
Obviously in order to compare classical with quantum predictions one has first
to smooth the step potential. Thus for the analysis of dynamical localization 
we consider the potential 
\be
\label{newmodel}
V(q) = Si((a+q)/\sigma) + Si((a-q)/\sigma)
\ee
where $Si(q) = \int_{0}^{q}\frac{\sin(t)}{t}dt$ is the sine integral function. 
For $\sigma \rightarrow 0$ we recover the step-like potential.
The classical force associated to this potential is 
%$F(q) = \frac{\sigma\sin[(a-q)/\sigma]}{a-q}- \frac{\sigma\sin[(a+q)/\sigma]}{a+q}$ 
$F(q) = \frac{1}{a-q}\sin\frac{a-q}{\sigma}- \frac{1}{a+q}\sin\frac{a+q}{\sigma}$ 
(Dirac delta functions in the limit $\sigma \rightarrow 0$). With this potential we get 
a well defined classical limit for any finite $\sigma$. We have computed the classical $P(p,t)$
by evolving the classical equation of motion for $10^6$ different random initial conditions 
with zero momentum $p = 0$ and uniformly distributed position along the interval $(-\pi,\pi)$.
We have found that $P(p,t) \sim 2Dt/p^2$ for $|p| < c(\sigma) \sqrt{2Dt}$ where $D \sim 1/{2\sigma}$ and
 $c(\sigma)$ increases as $\sigma$ decreases. Outside this region $P(p,t)$ resembles that of a standard
 diffusion process.  Thus for sufficiently small $p$ and $t$ the diffusion is anomalous and then gradually    
becomes normal.
% We add a smooth perturbation $V_{per}=8\cos(q)$ to 
%Eq. (\ref{potential}) in order to have a classical chaotic limit and thus 
%facilitate the analysis of dynamical localization.
In Fig. 6 we plot the quantum and classical 
variance $\langle p^2(t) \rangle$ of the density of probability $P(p,t)$ 
($p=\hbar k$ for the quantum case) measured at time $t$.
Calculations were carried out for the differentiable potential of Eq.(\ref{newmodel}).
It is observed that, as expected, for short times both classical and quantum results coincide. 
However, after a certain breaking time $t_b \propto D \approx 100$ 
the quantum particle still diffuses but a slower rate than the classical one thus suggesting that 
diffusion is weaken by quantum interference effects. We relate this new region of weak dynamical 
localization the the effect of the classical singularity on the quantum dynamics.

For even longer times $t_c \sim 20000$ related to the
crossover to normal classical diffusion 
the quantum particle tends to localize in momentum space as a consequence of standard dynamical localization effects. 
The latter regime is due to the underlying smooth nature of the potential investigated.
%This is reinforced by the fact the breaking time $t_b$ does not depend too much on $\sigma$ 
%{\bf (why?)} but the transition to classical almost normal diffusion and 
%quantum localization observed for longer times is strongly dependent on what $\sigma$ is
%used. 
Thus in order to observe genuine quantum 
effects associated to classical singularities the value 
 of $\sigma$ must be such that $t_b \ll t_c$. We remark that this condition 
 should be met by any experiment aiming to confirm the results reported in this letter.

% We recall that unlike the 
% standard kicked rotor the quantum diffusion never stops so 
%the transition to localization is only partial. This is consistent with 
% the eigenvector analysis below. 

For the sake of completeness we have also 
investigated the specific form of the quantum $P(k,t)$ as a function of $k$ (from 
now on we switch back to our original potential Eq.(\ref{potential})). 
In the context of a disordered conductor it has been 
reported that \cite{huck} at the AT the quantum diffusion is anomalous.  For 
 time scales large enough the density of
 probability has a power-law form with the exponent depending on the multifractal
dimension $D_2$. 
Similar results has also been recently obtained in momentum space
 for a kicked rotor with a 
logarithmic singularity \cite{ant9}. 
%In the latter case it is found that the overall effect of 
%the quantum corrections is also to suppress ({\bf or enhance?}) the classical 
%(anomalous) diffusion but at a rate much weaker than that in the case for smooth 
%potentials where after a certain time the quantum diffusion is fully suppressed. 
%
%
As expected, we have also observed power-law tails in our model in both
 the short and the long $t$ limit.  For
$|k|\gg t/\hbar\rho$ (with $\rho = 1/2\pi$ the spectral density) a best fit 
 estimates yields $P(k,t)\sim t^{0.84 \pm 0.01}
|k|^{-2}$ (see Fig 7), similar to the classical prediction. 
By contrast in the opposite limit though power-law tails have also been
 found the diffusion is slower in agreement with previous findings. 
%Numerical results suggest that  independently of the details of the classical potential. 

%We find that the power exponent
%$\gamma (t)$ is initially linear in $t$ (starting from $-2$ which is a result 
%of the power-law decay of the evolution matrix $U_{mn}\sim 1/|m-n|$)
%first, then tends to level off at around $-1.35$. 

%A1
%Thus for short time scales, unlike the case of a smooth potential, the
%quantum diffusion is indeed faster than the classical one. The reason for that is that
%the step-like singularity has no effect on the classical motion, but quantum 
%mechanically induces long-range power-law hopping which makes the diffusion faster. 
Another feature induced by the step-like singularity is the
 power-law localization of the eigenstates. In Fig. 8 is clearly observed that the 
eigenstates \cite{note} have power-law tails with an exponent around minus one. 
 This is in clear contrast with the exponential 
 localization observed in the kicked rotor with a
smooth potential. However is similar to the case of a potential with a $\log$ singularity
\cite{ant9}; not surprisingly in both cases the off-diagonal
 elements of the evolution operator present a similar power-law decay \cite{ever}. 
We mention there is no contradiction between this smooth power-law behavior and the
 multifractal features investigated previously. The point is that the multifractal 
 character of the eigenstates appears as strong fluctuations around the smooth power-law behavior above.
Roughly speaking we can say that smooth power-law localization is the precursor of multifractality.

%%%%%%%%%%%%%Fig 8%%%%%%%%%%%%%%%%%%%%%%%%%%%%%%%%%%%%%%%%%%%%%%%%%%%%%%%%%%%%%%%%%%%%%
\begin{figure}
\includegraphics[width=.95\columnwidth,clip]{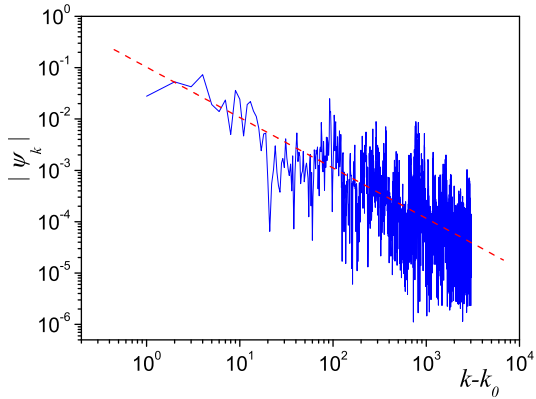}
\vspace{-.7cm}
\caption{(Color online) A typical eigenstate of the evolution matrix 
Eq.(\ref{uni})  ($T=1$, $\hbar=1$)
 with potential Eq. (\ref{potential}) ($v_0=\pi$ and $a=\pi/2$) plus a 
perturbation $V_{per}=8 \cos q$. The modulus of the eigenstate $|\psi_k|$
decays from its peak $k_0$ as a power-law with an exponent close to $-1$. 
Indeed the best fit (dashed line) corresponds to a slope $-0.98$.}
\label{figure8}
\end{figure}
%%%%%%%%%%%%%Fig 8%%%%%%%%%%%%%%%%%%%%%%%%%%%%%%%%%%%%%%%%%%%%%%%%%%%%%%%%%%%%%%%%%%%%%
Finally we discuss how the findings of this paper may be tested experimentally. 
The first direct experimental realization of the quantum kicked rotor with a smooth 
potential was reported by Raizen \cite{raizen} and coworkers in 1995. The experimental 
set up consisted of a dilute 
sample of ultracold atoms (typically Cesium or Sodium) `kicked' by a periodic standing
wave of near-resonant light that is pulsed on periodically in time to approximate a 
series of delta functions. The typical output of the experiment is the distribution of the 
atom momentum as a function of time. It is also possible to estimate the quantum breaking 
time signaling the beginning of quantum localization. For the case of a smooth periodic 
standing wave, dynamical localization was also detected experimentally in full agreement 
 with the theoretical predictions. We 
thus propose that the full $P(k,t)$, the accumulated probability  
or the variance represented in Fig. 6 may by accessible to experimental verification provided that
 the smooth periodic standing wave is replaced by the step-like wave studied in this paper. 
Obviously from a experimental point of view the
 step-like singularity is only approximated. Typically the experimental signal 
 is composed of a limited number of
Fourier components and consequently it is smooth on sufficiently small scales
 (as the potential discussed in connexion with dynamical localization). 
However, as we have shown in the previous section, 
we expect our results to hold in these 'almost' non-analytical potential at least up to 
a certain time scale related to the underlying smoothness of the potential.

\section{Conclusions}

In this paper we have studied a kicked rotator with a non-analytical step-like singularity.
It has been identified a region of parameters where the level statistics is exactly given 
by SP. These results are universal in the sense that they do not depend on the specific form of the 
potential but only on the presence of the classical singularity. The eigenfunctions have been shown to be 
multifractal but with a set of multifractal exponents $D_q$ different from the prediction 
of critical statistics. We have also conjectured, based on the numerical analysis and the 
comparison to other models with SP, that all systems described by SP have the same form of 
$D_q$ ($D_q =A/q+D_{\infty}$). Finally we have studied transport properties. It has been 
found that, unlike the standard kicked rotor,  dynamical localization slow down but 
does not stop quantum diffusion.
%This feature has been traced back to the fact that the 
%classical dynamics is not affected by  the step-like singularity. 
We have also discussed 
the possibility of experimental verification of our findings by using ultra cold atoms kicked 
by a standing wave with an approximated step-like form.      

\begin{acknowledgments}

AMG acknowledges financial support from a Marie Curie Outgoing Fellowship, contract MOIF-CT-2005-007300.  
JW is supported by DSTA Singapore under Project Agreement No. POD0001821.
\end{acknowledgments}


\begin{thebibliography}{9}
\bibitem{note} In the cylinder representation the eigenstates are obtained by 
               diagonalizing a non Unitary
               truncated evolution operator. In order to evaluate the error 
               involved in this approximation, we diagonalize both a truncated matrix of size $N$, 
               ${\cal U}^{(N)}$, and a larger truncated matrix, 
               say ${\cal U}^{(2N)}$, with ${\cal U}^{(N)}$ as its center. The truncation error
               is estimated to be $|({\cal U}^{(2N)}-\epsilon)\psi|$ where $\epsilon$ and $\psi$ are
               the eigenvalue and eigenstates of $\cal U$ respectively. In this way we find that for
               the range of volumes considered in this paper about one tenth of the total $N$ 
               eigenstates are not barely affected by the truncation approximation. 
               The eigenstate presented in Fig. \ref{figure8} is chosen from these one tenth.
\end{thebibliography}
\end{document}